\documentclass[11pt,oneside,onecolumn]{article}
\usepackage{algorithm2e}
\usepackage{booktabs}
\usepackage{graphicx}
\usepackage{verbatim}
\usepackage{enumitem}
\usepackage{url}

\DeclareGraphicsExtensions{.pdf,.png,.jpg}
\begin{document}
\title{An Improved Neighbourhood for the Traveling Tournament Problem}
\author{Glenn Langford\footnote{e-mail: glenn.langford@gmail.com} }
\date{July 3, 2010}
\maketitle

\begin{abstract}
The Traveling Tournament Problem (TTP) is a challenging combinatorial optimization problem that has attracted the interest of researchers around the world. This paper proposes an improved search neighbourhood for the TTP that has been tested in a simulated annealing context. The neighbourhood encompasses both feasible and infeasible schedules, and can be generated efficiently. For the largest TTP challenge problems with up to 40 teams, solutions found using this neighbourhood are the best currently known, and for smaller problems with 10 teams, three solutions found were subsequently proven optimal.
\end{abstract}
%
%
\section{Introduction}
The Traveling Tournament Problem (TTP) proposed in \cite{trick2001} requires scheduling a double round-robin tournament while satisfying a set of constraints. The best schedules also minimize the total travel distance of the tournament participants.

Similar to the Traveling Salesman Problem (TSP), the TTP requires searching for a shortest tour through a number of destinations. However, unlike the TSP the TTP requires the creation of interlocking travel schedules across multiple participants, with tight constraints on feasible solutions. 

Effective neighbourhood designs have been developed for the TTP using tabu search and simulated annealing in \cite{digaspero2007} and \cite{vanhentenryck2006}. However, it is desirable to have a neighbourhood design that affects fewer games at each transition, especially as the number of teams increases. In a markov chain context, we might expect such a neighbourhood to allow a sequence to reach quasi-equilibrium faster. In addition, a larger neighbourhood might also help to ensure that a given search procedure is not trapped in a local minimum.

The rest of this paper is organized as follows: a brief description is given of some known neighbourhood structures encompassing feasible and infeasible schedules. This background discussion is a slightly different presentation of the techniques used in \cite{vanhentenryck2006}. A proposal for an improved neighbourhood is outlined, and its properties examined. Finally, a summary of experimental results using a simulated annealing approach is presented.
%

\section{Background of the Traveling Tournament Problem}
In a TTP schedule, each of \(n\) participants or teams must play every other opponent twice; once at the team's home location, and once at the opponent's location. Thus, there are \(2n - 2\) games in each team's schedule. TTP challenge problems specify a distance matrix \(D\) defining the travel distance between all teams. Traveling between the locations of two teams \(t_i\) and \(t_j\) incurs a travel cost of \(D_{ij}==D_{ji}\). By definition, every team must return home at the end of the schedule; if \(t_i\) is away at the location of some team \(t_x\) in the last game of the schedule, \(D_{xi}\) is added to the total travel distance for \(t_i\).

TTP schedules commonly have two significant ``soft'' constraints, beyond the double round-robin requirement. A schedule satisfying these constraints is said to be \emph{feasible}, while a schedule which contains one or more violations to these constraints is \emph{infeasible}. The \emph{At Most} constraint requires that no more than three consecutive home or away games are allowed for any team. The \emph{No Repeat} constraint requires that no team play the same opponent in consecutive rounds. (The mirror constraint is not considered in this paper, which is discussed in \cite{urrutia2007}).

Figure \ref{gal10-4535} shows an example (shortest) schedule for the Galaxy 10 problem, discovered using the techniques described later.
\begin{figure}
\include{gal10-4535}
\caption{A tournament schedule for the Galaxy 10 problem comprising \(2*10-2=18\) rounds which satisfy the At Most and No Repeat constraints. The ``@'' symbol designates an away game, and if \(t_i\) plays away to \(t_j\) in a round, then by definition \(t_j\) is at home to \(t_i\) in that same round.}
\label{gal10-4535}
\end{figure}

\subsection{Basic Neighbourhood}
Following the definition and syntax outlined in \cite{vanhentenryck2006}, a basic TTP neighbourhood is reviewed on which later work is based.

A \emph{move} is a transformation which can be applied to one schedule to reach one or more other schedules. Schedules may be feasible or infeasible, but are always complete double round-robin tournaments at the completion of a move. Multiple types of moves may be defined. The union of schedules reachable in a single move, across all types of moves, defines the \emph{neighbourhood}.

Given a TTP with n teams, (n even) with a set of teams \(1 \leq t_i \leq n\) and rounds \(0 \leq r_i < 2n-2\),
a neighbourhood can be constructed from the following set of moves:
\begin{enumerate}
\item
\(swapHomes(t_i, t_j)\) reverses the home/away status whenever \(t_i\) plays \(t_j\). This must occur exactly twice in the schedule for \(t_i\) and \(t_j\) in the same rounds. All other teams remain unchanged.

\item
\(swapRounds(r_k, r_l)\) swaps all games in round \(r_k\) with \(r_l\) across all teams. 

\item
\(swapTeams(t_i, t_j)\) swaps the schedule of teams \(t_i\) and \(t_j\), except when the two teams play each other. The schedules of all the other teams are also updated to reflect the swap. A variant of \(swapTeams\) combines \(swapTeams(t_i, t_j)\) with \(swapHomes(t_i, t_j)\) which produces a new schedule which is violation neutral with respect to the previous schedule (it neither increases nor decreases the number of constraint violations in the schedule).

\item
\(partialSwapRounds(t_i, r_k, r_l)\) starts with a swap of the games in rounds \(r_k\) and \(r_l\) for team \(t_i\). Suppose \(t_i\)'s opponents in these two games are \(t_x\) and \(t_y\); we then continue to swap the same rounds for \(t_x\) and \(t_y\), making note of their opponents in the swapped rounds. The process continues until all opponents in any swapped rounds have had the same swap performed; this ensures that a double round-robin tournament is produced that is consistent.

\item
\(partialSwapTeams(t_i, t_j, r_k)\) starts with a swap of the game between \(t_i\) and \(t_j\) in round \(r_k\). Suppose that \(t_i\) swapped away a home game against \(t_x\) in that round, and received a home game with \(t_y\) in return. From the perspective of \(t_i\), the schedule now has a duplicate home game with \(t_y\): one in round \(r_k\), and another in some other round, \(r_l\). The swap of games between \(t_i\) and \(t_j\) then continues in round \(r_l\) to trade away the duplicated game, and the swapping procedure continues. Swapping stops only when a home game with \(t_x\) is received by \(t_i\) in a swap. This procedure ensures that a valid double round-robin tournament is preserved, with (hopefully) fewer changes than \(swapTeams(t_i, t_j)\).
\end{enumerate}
\begin{table} 
	\centering
	\begin{tabular}{cccc}
	\toprule
	& \multicolumn{2}{ c}{Team Schedules} \\
	\cmidrule(r){2-3}
	Round & A & B & Swap sequence \\
	\midrule
	0 & B & @A &  \\
	1 & C & @H & 1 \\
	2 & D & @C &  \\
	3 & E & @F & 9 \\
	4 & F & @G & 5 \\
	5 & G & @E & 3 \\
	6 & H & @D & 7 \\
	7 & @B & A &  \\
	8 & @C & D &  \\
	9 & @D & E &  8\\
	10 & @E & F & 4 \\
	11 & @F & C & 10 \\
	12 & @G & H & 6\\
	13 & @H & G & 2 \\
	\bottomrule
	\end{tabular}
	\caption{An example schedule fragment (highly infeasible!) for two teams, A and B, part of an 8 team schedule. The schedules for the other six teams (C through H) are elided. The games shown are prior to executing \(partialSwapTeams(A, B, 1)\). The first game swap would take place in round 1, where team A would receive '@H'. The next game to be swapped is thus round 13, where A has a duplicate '@H' game. The home game against team C would be returned to team A after 10 swaps (this game was lost in the first swap), and the \(partialSwapTeams\) swap sequence is then complete.}\label{tab:pst} 
\end{table}

To illustrate the machinery in \(partialSwapTeams\), Table \ref{tab:pst} shows a progression of swaps in a sample schedule for the move \(partialSwapTeams(A, B, 1)\). While round 1 is the starting point shown here, note that the resulting schedule would be the same if any round in that move's swap sequence was selected as the starting round. This is true in general for \(partialSwapTeams\), and as a result the \(partialSwapTeams\) effective neighbourhood size is substantially smaller than the number of \(partialSwapTeams\) moves for a given schedule.

\section{An Improved Neighbourhood}
This section outlines a new strategy for extending \(partialSwapTeams\), creating a larger accessible neighbourhood with improved acceptance rates.

The strategy is based on preconditioning a schedule to reduce the number of games swapped by the \(partialSwapTeams\) procedure. For example, Table \ref{tab:pst2} illustrates the scope of change similar to Table \ref{tab:pst}, but with a perturbation first applied to the schedule in the form of a \(swapHomes\) move. This perturbation supposes we are prescient enough to precondition schedules with some chosen \(swapHomes\) move prior to executing \(partialSwapTeams\). 

Simulating the same \(partialSwapTeams(A, B, 1)\) move as before on the altered schedule, only 5 rounds are involved in exchanges between A and B rather than 10. Of course, this gain was not achieved without cost, since the \(swapHomes\) move also disturbed the schedule. This can be partially reversed if we simply repeat the same \(swapHomes\) move as an additional step.

\begin{table} 
	\centering
	\begin{tabular}{cccc}
	\toprule
	& \multicolumn{2}{ c}{Team Schedules} \\
	\cmidrule(r){2-3}
	Round & A & B & Swap sequence \\
	\midrule
	0 & B & @A &  \\
	1 & C & @H & 1 \\
	2 & D & @C &  \\
	3 & E & @F & 4 \\
	4 & F & @G &  \\
	5 & G & \fbox{ E} & 3 \\
	6 & H & @D &  \\
	7 & @B & A &  \\
	8 & @C & D &  \\
	9 & @D & \fbox{@E} &  *\\
	10 & @E & F &  \\
	11 & @F & C & 5 \\
	12 & @G & H &  \\
	13 & @H & G & 2 \\
	\bottomrule
	\end{tabular}
	\caption{A schedule fragment for two teams, A and B, part of an 8 team schedule. The schedule is derived from 
	Table \ref{tab:pst} with the home/away status reversed for team E in the schedule for team B. In other words, \(swapHomes(B, E)\) has been executed as a preconditioning step. If we then execute \(partialSwapTeams(A, B, 1)\) as before, the number of swapped rounds is reduced to 5 from 10. Round 9 is shown in the swap sequence with a '*' since it has been changed; however, if \(swapHomes(B, E)\) is repeated after \(partialSwapTeams(A, B, 1)\) then round 9 is returned to its original game assignments.}\label{tab:pst2} 
\end{table}

\subsection{Lookahead Partial Swap Teams}
Generalizing from this example, consider a new compound move called \(lookAheadPartialSwapTeams\), or L-PST, sketched in Algorithm \ref{L-PST}. This move is constructed by preconditioning a schedule with \(swapHomes\) before executing the usual \(partialSwapTeams\) procedure.

\begin{algorithm}
	\SetAlgoLined    
	\KwData{ \( (t_i, t_j, r) \) }
	select look ahead opponent \(t_l\)\;
	\(swapHomes( t_j, t_l ) \)\;
	\(partialSwapTeams( t_i, t_j, r ) \)\;
	\(swapHomes( t_j, t_l ) \)\;
\caption{Sketch of basic Lookahead Partial Swap Teams}\label{L-PST}
\end{algorithm}

The value of L-PST depends on the selection of a \emph{look ahead opponent} \(t_l\) that reduces the number of game swaps and overcomes the added cost of \(swapHomes\). 

The search for a look ahead opponent starts by simulating the generic \(partialSwapTeams(t_i, t_j, r)\) procedure, collecting all of the opponents of \(t_j\) into a list, called a \emph{swaplist}. For example, looking at the swap sequence shown in Table \ref{tab:pst}, the swaplist constructed from the opponents of team B is:
\begin{equation}
  \{@H, G, @E, F, @G, H, @D, E, @F, C\}   
\end{equation}

To identify candidates for the look ahead opponent, a swaplist is scanned for duplicate teams with a sufficiently large gap between them; any team that occurs twice in the swaplist is a candidate look ahead opponent. The gap size represents the benefit of the \(swapHomes\) preconditioning step. For example, team H is a good candidate since there is a gap of 5 between @H and H. The same gap length exists for teams E and F. 
\begin{equation}
  \{ \overbrace{@H, G, @E, F, @G, H}^5, @D, E, @F, C\}
\end{equation}

Note that choosing either H, E, or F as the look ahead opponent generates three different resulting schedules, increasing the size of the neighbourhood that can be reached in one move.\footnote{The opponents in the swaplist that occur within the gap framed by the look ahead opponent are skipped by the \(partialSwapTeams\) portion of the L-PST procedure, therefore selecting a different look ahead opponent will result in a different schedule.} These three new schedules are all achieved with fewer total game swaps than the original \(partialSwapTeams\) procedure. In each case, the number of expected game swaps is 5, which is predicted by the length of the swaplist less the look ahead opponent gap size. In practice, selection of a look ahead opponent is based on finding the largest gap in the swaplist; ties can be resolved by coin tossing.

Of course, it is possible that no duplicate opponents will be found in a given L-PST swaplist, and therefore the preconditioning strategy is not available. In this case, the usual \(partialSwapTeams\) procedure is followed. Algorithm \ref{L-PST-2} shows a version of L-PST with this feature.
\begin{algorithm}
	\SetAlgoLined
	\KwData{ \( (t_i, t_j, r) \) }
	simulate generic \(partialSwapTeams( t_i, t_j, r ) \) procedure to produce a swaplist for \(t_j\)\;
	select look ahead opponent \(t_l\)\;
	\eIf{ \(t_l\) available }{
		\(swapHomes( t_j, t_l ) \)\;
		\(partialSwapTeams( t_i, t_j, r ) \)\;
		\(swapHomes( t_j, t_l ) \)\;
		}{
		\(partialSwapTeams( t_i, t_j, r ) \)\;
	}
\caption{Basic Lookahead Partial Swap Teams algorithm}\label{L-PST-2}
\end{algorithm}

\subsection{Early Exit Strategy}
It is possible to further curtail the number of game swaps with a strategy termed \emph{early exit}. Recall that the generic \(partialSwapTeams( t_i, t_j, r )\) procedure first swaps a game between \(t_i\) and \(t_j\) in round \(r\). Team \(t_i\) loses some game against an opponent \(t_x\) with this first swap. Ordinarily, if this round was \(t_i\)'s home game against \(t_x\) (for example), swapping would continue until a home game against \(t_x\) is received from \(t_j\). However, it is possible to stop earlier, if an \emph{away} game against \(t_x\) is received instead. Stopping at this point leaves the schedule in a state which is not double round-robin; however, it can be repaired by noticing that \(t_i\) has two away games against \(t_x\), and flipping one of the two games at random to be a home game restores the double round-robin property for \(t_i\). \(t_j\) has a similar problem, with two home games against \(t_x\), one of which must be flipped to an away game. Of course, the procedure must also correct the two corresponding games for \(t_x\).

In general, the early exit strategy can be followed in a \(partialSwapTeams( t_i, t_j, r )\) move when \(t_i\) regains a game which has the original round \(r\) opponent, but with the wrong home/away status.

\subsection{Combining Look Ahead with Early Exit}
The lookahead and early exit strategies can be combined to reduce the number of game swaps. For example, suppose that the L-PST simulation step in Algorithm \ref{L-PST-2} produced the following swaplist: 
\begin{equation}
   \{ \overbrace{A, B, C, D, E, @A}^5, @F, B, C, D, E, F\}
\end{equation}
Choosing B, C, D, or E as the look ahead opponent, each with a gap of 6, is certainly a big improvement. But it is also possible to choose A (with a gap of 5), and exit early with @F rather than continuing with 5 additional swaps to get F. In general, to choose the look ahead opponent penalties are provisionally added to the \(swapHomes\) preconditioning step and the early exit (when applicable), and the shortest overall path is constructed to either the final opponent in the swaplist, or an earlier match against the same opponent.

The complete version of L-PST is summarized in Algorithm \ref{L-PST-3}.

\begin{algorithm}
	\SetAlgoLined
	\KwData{ \( (t_i, t_j, r) \) }
	simulate generic \(partialSwapTeams( t_i, t_j, r ) \) to create a swaplist for \(t_j\)\;
	select look ahead opponent \(t_l\)\;
	\eIf{ \(t_l\) available }{
		\(swapHomes( t_j, t_l ) \)\;
		\(partialSwapTeams( t_i, t_j, r ) \)\;
		\(swapHomes( t_j, t_l ) \)\;
		}{
		\(partialSwapTeams( t_i, t_j, r ) \)\;
	}
	\If{ early exit was performed }{
		repair schedules for \(t_i\), \(t_j\) and opponent to restore double round-robin property\;
	}
\caption{Lookahead Partial Swap Teams with optional early exit}\label{L-PST-3}
\end{algorithm}

\section{Experiments} 
The shortest known schedules for the TTP challenge problems NL10, CIRC10, and Galaxy10 were all discovered by the author using Lookahead PST. These schedules were recently proven optimal in \cite{uthus2010}. Note that the optimal schedule was not found in every run; problems of this small size execute relatively quickly and were used for test purposes with dozens of runs for each. 

L-PST was also used to find all of the current best known schedules for the larger Galaxy problems. The results for the largest problems are summarized in Table \ref{tab:results}.

All results were obtained using a single desktop computer, using a simplified version of the TTSA algorithm without reheats \cite{vanhentenryck2006}, starting from random initial schedules. Other aspects of the TTSA algorithm including the objective function were also modified. The proposal distribution used in TTSA is not described in \cite{vanhentenryck2006}; as an indication of the relative importance of L-PST, in the largest challenge problem (Galaxy with n=40) the initial proposal distribution used suggests L-PST for approximately 66 \% of all move attempts.

\begin{table} 
	\centering
	\begin{tabular}{cccc}
	\toprule
	Problem size & Previous best & L-PST best & Improvement \\
	\midrule
	36 & 207,117 & 177,090 & 14.5\%\\
	38 & 253,279 & 214,546 & 15.3\% \\
	40 & 304,689 & 258,899 & 15.0\% \\
	\bottomrule
	\end{tabular}
	\caption{Summary of results for the largest Galaxy challenge problems.}\label{tab:results} 
\end{table}

\section{Conclusion}
An improved search neighbourhood for the Traveling Tournament Problem is proposed which uses a derivative of the \(partialSwapTeams\) move. New mechanisms were outlined which identify shortcuts to reduce the number of game swaps. With a reduced number of game swaps, on average, the goal is to avoid becoming trapped in a local minimum during a search procedure. The proposed Lookahead Partial Swap Teams or L-PST algorithm used in a simulated annealing context appears to be effective in finding optimal schedules for problems up to 10 teams, and very good, short distance schedules for a wide range of larger challenge problems. Whether the solutions to larger problems are close to optimality or not remains to be proven.

%
%

%
%
\end{document}